# Workflow for Safe-AI

Suzana Veljanovska
Institute of Embedded Systems
ZHAW School of Engineering
Winterthur, Switzerland
veln@zhaw.ch

Hans Dermot Doran
Institute of Embedded Systems
ZHAW School of Engineering
Winterthur, Switzerland
donn@zhaw.ch

*Abstract*—The development and deployment of safe and dependable AI models is crucial in applications where functional safety is a key concern. Given the rapid advancement in AI research and the relative novelty of the safe-AI domain, there is an increasing need for a workflow that balances stability with adaptability. This work proposes a transparent, complete, yet flexible and lightweight workflow that highlights both reliability and qualifiability. The core idea is that the workflow must be qualifiable, which demands the use of qualified tools. Tool qualification is a resource-intensive process, both in terms of time and cost. We therefore place value on a lightweight workflow featuring a minimal number of tools with limited features. The workflow is built upon an extended ONNX model description allowing for validation of AI algorithms from their generation to runtime deployment. This validation is essential to ensure that models are validated before being reliably deployed across different runtimes, particularly in mixed-criticality systems.

## I. Introduction

Developing functional safety (FS) systems requires achieving exceptionally low failure probabilities in both hardware and software. These systems typically adhere to the sample-compute-actuate paradigm, ensuring that parameters used alongside sampled values to compute actuator outputs are meaningful, well-understood, and justifiable and qualifiable.

In contrast, AI workflows operate differently. The parameters of the AI algorithms, particularly in deep neural networks, are derived from training processes that often rely on data from uncertain sources. Because the semantic origins of these parameters are not clearly defined, they cannot be readily justified to independent authorities. Furthermore, with the recent advancement in the AI field, current models contain an immense number of weights—ranging from millions to even a billion intended to capture the features learned from training data. Despite the large number of weights, their connection to the physical environment of the problem is highly difficult to isolate and interpret. While it is commonly understood that this challenge forms the foundation of the research domain of explainable AI [1], determinism as opposed to explainability is critical to the safe AI.

For highly complex and context-dependent applications, such as the automotive domain, there is an increasing tendency to validate AI classifier outputs using other AI-based approaches, such as classical expert systems [2]. However, in many applications where contextual dependency is lower, we argue that the number of reliably identifiable classes is relatively small compared to the total number of classes an AI classifier might recognize [3].

Ensuring that AI classifiers are safe, dependable and reliable, particularly deep neural networks require a universally accepted workflow. We assert that the field of functional safety represents the most rigorous standard for workflow stringency. Consequently, the acceptance of AI workflows depends on their alignment with established functional safety practices [4], which themselves originate from best practices in general engineering.

Ensuring safety and dependability of AI execution and deployment is crucial, particularly in high-criticality applications. This paper presents a qualifiable AI workflow featuring minimal tool dependencies, focusing on the ONNX format for model representation [5]. The proposed workflow includes model validation at different stages to ensure integrity before deployment. Since tool qualification follows a rigorous process, making it both resource and time intensive [6], the model will be trained utilizing unqualified tools. Therefore, a validator is necessary to ensure the validity of the outputs generated by these tools. We describe two validators in the workflow, the first verifies architectural consistency between pre-trained and trained models, while the second ensures that the partitioned models with added reliability features, recombined, retain their original architecture. The architecture validators are implemented in Python.

## II. Motivation and Previous Work

The current lack of a well-defined workflow is the motivator for this paper considering the essential features of an AI workflow designed for functional safety, dependability, or reliability. Our previous work captures the use of the single protected channel pattern (SPCP) [6] as a method for reliably validating CNN outputs (Figure 1) for safe and dependable execution [7].

Our use case is a shape recognition use case for anomalous shapes detection in the industrial domain. We utilize AlexNet as a CNN for this purpose as a widely utilized model in object classification tasks that reaches relatively high accuracy. The training process is performed using a dataset containing various objects. For the shape validation we use Symbolic Aggregate Approximation (SAX) [8][9] due to its deterministic nature. SAX is a dimension reduction algorithm that reduces time-series data to a string while preserving the crucial features of the shape and allowing inexpensive comparison with other strings. The time-series of a shape is generated by the distance from the centroid of the shape to its edges.

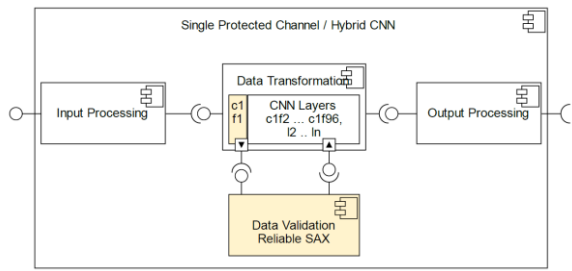

FIGURE 1 SINGLE PROTECTED CHANNEL PATTERN FOR CNN OUTPUT VALIDATION [3]

Since the first task of the validation algorithm is a shape contour extraction, it expects a binarized image as input with clearly defined edges. For this purpose, an edge extraction on the input image using Sobel filter along the x and y-axes has been applied. For lowering the computation, a hybrid approach has been implemented by incorporating this filtering in the CNN and replacing the first of the 96 channels from the first convolutional layer with the Sobel kernels. We have shown [10] that the kernel replacement does not negatively affect the CNN's accuracy as it remains the same as the original model although the confusion matrices slightly differ. This first (kernel/channel) can be executed reliably, thus instantiating a reliable execution path within the model by which non-reliably derived results of the entire model may be qualified.

To guarantee dependable input for the qualifier, the convolution process utilizing the specific Sobel filter is executed redundantly. Since convolution involves multiplications and additions, we overload these operators and execute the operations redundantly. Instead of returning a single value, the overloaded functions return both the qualifier value and the operation's output, verifying whether the redundant computation yields identical results. The complete reliable convolution is detailed in [10].

III. WORKFLOW FOR SAFE-AI DEPLOYMENT

The general idea for the workflow is that for it to be considered safe it needs to be qualifiable which necessitates the use of qualifiable tools. Tool qualification requires effort and expense. In order to minimize the expense , this work focuses on implementation of a lightweight workflow utilizing as few as possible tools with limited features.

The line of thought is the following, instead of creating and training the model with the well-known tools such as PyTorch or Tensorflow, the model is directly built as onnx graph and trained as onnx model. We prove that there is no significant difference in the accuracy when training the same model in Tensorflow and ONNX on the same dataset. The creation and training of the model utilizing the same tool as the rest of the workflow allows to keep the traceability of the model from the initial creation and all the way down to deployment.

The diagram in the Figure 2 has been designed to describe the workflow based on the previous considerations. Yellow/shaded regions refer to qualifiable tools. The creation and definition of the model is directly specified in ONNX format (.onnx file A). This file can be used to train the algorithm utilizing a standard tool and produce an output ONNX representation (.onnx file B). We first validate that both files A and B have the same architecture, this ensures hat only the trainable parameters in the training process have been modified and the non-trainable parameters as well as the general topology of the network remain unchanged. This ensures the algorithm's integrity as well as the validity of the training process. The validator produces a qualified ONNX representation encoded in .onnx file C. The qualified file alongside with the file containing the reliability attributes serves as input to the Model Partitioner which generates two files – one for runtime execution on a reliable platform (.onnx file D) and one for runtime execution on non-reliable platforms (.onnx file E). The partitioner also accepts an attribute file with directions as to how the model is to be partitioned. The Architecture Validator must be able to validate that the two runtime files when recombined produce the intended CNN architecture.

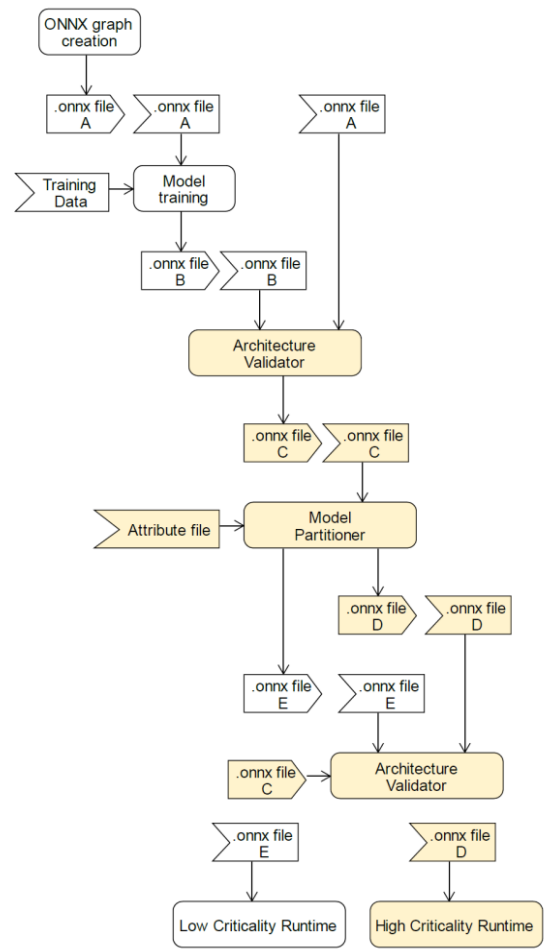

FIGURE 2 WORKFLOW FOR SAFE-AI DEPLOYMENT [11]

## IV. ARCHITECTURE VALIDATORS

The validation process ensures the integrity and consistency of AI models by employing two distinct but complementary validators.

The first architecture validator in the workflow focuses on verifying that the architecture of a pre-trained model remains unchanged after training, with modifications strictly limited to weight updates. It begins by analyzing the pre-trained model and assigning it as the reference, extracting the metadata, and generating a human-readable report outlining its algorithmic structure, inputs-outputs mapping of the layers, non-trainable hyperparameters, and the overall layers topology of the algorithm.

The trained model is subjected to the same analysis, with its extracted attributes compared against the reference. If differences are detected such as inconsistencies in layers, hyperparameters, or input-output mappings between the layers an error report is generated detailing these inconsistencies. If no structural changes are found, the validator produces a validated ONNX representation with embedded metadata indicating that the file had been successfully validated. This step ensures that the training processes did not modify the architecture of the algorithm, preserving the model's validity and traceability.

The second architecture validator is responsible for validating partitioned files, ensuring that when combined, they accurately reconstruct the original trained model while incorporating necessary reliability-enhancing features.

The validator first incorporates the separate files into a single report by merging them based on the attribute specifications from the partitioning process and verifies that layer connections remain aligned across files. Each layer includes metadata specifying its allocation, allowing the validator to reconstruct the complete model. Structural comparisons between the reconstructed model and the original trained version are then applied to detect any deviations between the models.

Safety functions, such as redundancy layers or external validation functions are identified and documented in the validation report. If the partitioning introduces unauthorized modifications, a discrepancy report is generated stating those deviations, otherwise, validated files are produced for the high-criticality execution. This validator ensures that the partitioning maintains the core architecture while enabling necessary enhancements for system reliability.

## V. DEVELOPMENT PROCESS

The traceability and qualifiability of the workflow facilitates adherence to the V-model, a well-established development methodology that ensures verification and validation at each stage of the process. This process enhances system integrity and reliability by systematically aligning the design and testing phases.

The left side of the V-model represents the system development consisted of requirements specification, system, architecture and model design while the right side corresponds to the verification steps, including unit, system and architecture testing and the final validation (Figure 3). Each of the testing phases correspond to a design phase on the left-hand side, ensuring that every sequence is verified systematically. Both sides of the model are bridged by code generation, forming a structured approach to build and verify the software at each stage. Finally, the coding phase at the base of the V-model incorporates the designs into actual software components.

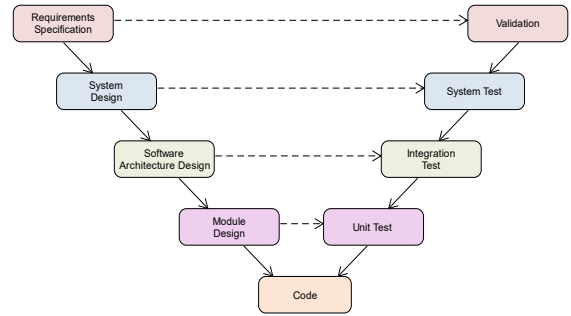

**FIGURE 3 V-MODEL DEVELOPMENT**

The first phase of development, requirements specification, involves defining the key parameters and specifications necessary for architecture validation. This includes outlining safety-relevant functions and features, establishing how the system partitions files for reliable and non-reliable execution, and selecting the programming language to be used for implementation. A validation step is conducted to ensure the defined specifications align with system requirements and safety constraints. The test cases considered at this stage will be a small subset of the system tests verifying the functional correctness of the system.

The system design phase focuses on defining the logical structure of the system, detailing the design of the validators and their associated functions. This phase specifies each validator's tasks, including expected inputs, processing logic, and generated outputs. The first validator inspects metadata and layer attributes, producing detailed diagnostic reports, while the second validator evaluates partitioned files and verifies the system's algorithmic architecture, ensuring compliance with safety constraints. The system testing phase involves constructing and analyzing various test scenarios and inputs for the system and defining the expected outcomes. This means generation of example input files and testing how the system behaves and if it produces the expected outputs.

The system architecture design phase provides a deeper focus on the internal structure of the system, emphasizing the interconnections between its individual components rather than treating the system as a singular entity. This phase defines the chains – function flows for the validators. This includes designing a detailed diagrams with the inputs and outputs described in higher details. It defines the lower-level logic of the separate functions on which the validators are built. The testing phase includes testing groups of units by sequentially adding units to the flow and verifying the outputs.

The module design phase, before the code design phase focuses on defining the functional modules that comprise the system's architecture. At this stage, each module is broken down into specific functions. Test cases are developed for each function, targeting edge cases, boundary conditions, and critical scenarios to validate robustness. Python's built-in testing capabilities, such as unittest and pytest, provide a structured and efficient means of implementing these tests, making unit testing more accessible and maintainable.

The code generation phase follows module specification and incorporates software safety and maintainability best practices. This includes ensuring the generated code remains clean, readable, and transparent, while avoiding critical structures such as unbounded recursion and deep nesting, both of which increase complexity and affect the traceability of the code which results in difficulties and challenges in the testing phases. By maintaining a structured, well-documented approach, the system remains transparent, verifiable.

## VI. Conclusion

This work presents a hybrid CNN model architecture for reliable and non-reliable execution. We maintain low computational cost by insertion of reliable features in the model where necessary. We incorporate the qualifier function that produces deterministic outputs to qualify safety-critical output classifications from the CNN. We then present a structured AI workflow with validation mechanisms ensuring safe and dependable deployment an execution. By leveraging ONNX as a neutral model representation, we minimize tool dependencies and enhance model transparency. The proposed validators ensure architectural integrity, enabling reliable AI deployment in safety-critical environments. We find this approach applicable in the applications where functional safety is the key factor and useful to mixed-criticality systems.